\documentclass[12pt]{article}
\usepackage{rotating}
\usepackage[T1]{fontenc}
\usepackage[utf8]{inputenc}
\usepackage{epsfig,amssymb}
\usepackage{ulem}
\input{epsf}

\def\laq{\raise 0.4 ex \hbox{$<$}\kern -0.8 em\lower 0.62 ex\hbox{$\sim$}}
\def\gaq{\raise 0.4 ex \hbox{$>$}\kern -0.7 em\lower 0.62 ex\hbox{$\sim$}}

\def\beq{\begin{equation}}
\def\eeq{\end{equation}}
\def\beqa{\begin{eqnarray}}
\def\eeqa{\end{eqnarray}}

 \def\frac#1#2{{\textstyle{{#1}\over {#2}}}}
 \def\lsim{\mathrel{\rlap{\lower4pt\hbox{\hskip1pt$\sim$}}
    \raise1pt\hbox{$<$}}} \def\gsim{\mathrel{\rlap{\lower4pt\hbox{\hskip1pt$\sim$}}
    \raise1pt\hbox{$>$}}}
\def\sqr#1#2{{\vcenter{\vbox{\hrule height.#2pt
         \hbox{\vrule width.#2pt height#1pt \kern#1pt
         \vrule width.#2pt}
         \hrule height.#2pt}}}}

 \def\frac#1#2{{\textstyle{{#1}\over
{#2}}}} 
\def\lsim{\mathrel{\rlap{\lower4pt\hbox{\hskip1pt$\sim$}}
\raise1pt\hbox{$<$}}}
\def\gsim{\mathrel{\rlap{\lower4pt\hbox{\hskip1pt$\sim$}}
\raise1pt\hbox{$>$}}} \def\sqr#1#2{{\vcenter{\vbox{\hrule height.#2pt
\hbox{\vrule width.#2pt height#1pt \kern#1pt \vrule width.#2pt} \hrule
height.#2pt}}}}

\def\beq{\begin{equation}} \def\eeq{\end{equation}}
\def\beqa{\begin{eqnarray}} \def\eeqa{\end{eqnarray}}

\def\gappeq{\mathrel{\rlap {\raise.5ex\hbox{$>$}} {\lower.5ex\hbox{$\sim$}}}}

\def\lappeq{\mathrel{\rlap{\raise.5ex\hbox{$<$}}
{\lower.5ex\hbox{$\sim$}}}}

\usepackage{color}

\begin{document}
\pagestyle{plain}

\begin{flushright}
May 2026
\end{flushright}
\vspace{15mm}

\begin{center}

{\Large\bf On an Airborne Proton Accelerator for Enhancing Cloud Formation or Inducing their Precipitation}

\vspace*{1.0cm}

Orfeu Bertolami$^{1,2}$ \\
\vspace*{0.5cm}
{$^{1}$ Departamento de F\'{\i}sica e Astronomia, Faculdade de Ci\^encias,
Universidade do Porto, \\
Rua do Campo Alegre s/n, 4169-007 Porto, Portugal}\\

{$^{2}$ Centro de F\'{\i}sica das Universidades do  Minho e do Porto,
Rua do Campo Alegre s/n, 4169-007 Porto, Portugal}\\

\vspace*{2.0cm}
\end{center}

\begin{abstract}

\noindent
We argue that an airborne proton accelerator is an interesting tool for weather control. Following the findings of the CLOUD experiment at CERN, 
one expects that a beam of protons, likewise cosmic rays and other aerosols, can enhance the formation of low-altitude clouds, 
allow for tailor made cooling of overheated areas and induce the precipitation of high-altitude clouds that trap solar radiation reflected from the ground. 
The proton accelerator can also be used to mitigate droughts, 
regularise  precipitation and avoid that it takes place through large and harmful storms. 
\end{abstract}

\vfill
\noindent\underline{\hskip 140pt}\\[4pt]
\noindent
{E-mail address: orfeu.bertolami@fc.up.pt}

\newpage

\noindent
1. Low-altitude cumulus, stratus and stratocumulus-type clouds, found below 2000 meters, play an important role in reflecting back to space incoming short wavelength radiation and contributing positively to the overall Earth's albedo.  
On the other hand, high-altitude cirrus-type cold clouds, situated between 5000 to about 13000 metres, absorb outgoing long-wavelength radiation from Earth's surface and contribute to global warming likewise greenhouse gases. 
In this work we discuss the benefits of using an airborne  proton 
accelerator in order to seed the condensation of low-altitude clouds and argue that this very device can, operating with a higher beam flux, be used to induce the precipitation of high-altitude clouds. Therefore, an 
airborne accelerator can be a flexible instrument for weather management and control, a matter that was subject of generic discussions about putative technological developments long ago \cite{Neumann}.  
The proposed accelerator can be placed in a plane, Zeppelin or rocket, but not as we shall see, in a satellite or space station. The airborne accelerator is meant to work as an artificial source of cosmic rays so to enhance the formation of low-altitude clouds or to precipitate high-altitude clouds, so to mitigate 
global warming due to anthropogenic emission of greenhouse gases. 

Of course, the formation and precipitation of clouds involve complex processes. These include, besides the suitable atmospheric conditions such as horizontal movement of air, winds that transport energy from Earth's surface in the form of sensible and latent heat. The former takes place through conduction and convection. Of course, formation of clouds and their precipitation also involve latent heat. Clouds are formed when air holds as much water vapour till its saturation point. Saturation can take place in two ways. First, moisture accumulates by reaching the maximum volume that air can hold. The other process requires dropping the moisture filled air temperature, lowering in this way the amount of moisture air can hold. Thus, saturation can be reached either through evaporation or condensation. When saturation occurs, moisture turn into visible water droplets in the form of clouds and fog\footnote{Fog is defined as visible moisture that begins below a hundred meters or so and limit horizontal visibility up to a 1 km. More precisely, the optical depth of air varies in the range $10^{-2} \lsim \tau < 0.5$, where the lower limit corresponds to a very clear air and the upper limit to haze, meaning that the radiation intensity, $I_0$ travelling a distance, $L$, through the air falls as $I(L) = I_0 e^{-\tau}$.}, which differ from each other only through their altitude \cite{Mason}. 

As for precipitation, condensation by itself is not the direct cause of rain, snow, sleet or hail. Precipitation requires that the moisture in clouds get sufficiently heavy so to fall due to gravity. This can take place through two processes: i) Ice crystals and water droplets in cold clouds coexist and due to an imbalance of water vapour pressure, the water droplets get into ice crystals, which become heavy enough to fall; ii) Water droplets in warm clouds collide and exchange their electric charges. Droplets of unlike charge attract one another and merge, thereby growing until they have sufficient weight to fall \cite{Mason}.

Clouds can be formed through a diverse combination of factors, but at the most fundamental physical level the basic ingredients for the formation of clouds are water and dust. On Earth, clouds are composed primarily of water 
either in liquid or in solid state. The water vapour content of the atmosphere varies from almost vanishing to up about 4 percent. However, without ``dirty air",  that is, dust, there would be no clouds at all or only high-altitude ice clouds \cite{Nese}. 
Even the ``cleanest" air found 
on Earth contains about 1000 dust particles/$m^3$. Dust is needed for condensation nuclei, sites on which water vapour may either condense or deposit as a liquid or solid. Certain types and shapes of dust and salt particles, such as sea salts and clay, 
make suitable condensation nuclei. These airborne particles such as dust and salt particles are generically referred to as Cloud Condensation Nuclei (CCN), having typically $(10^{-3} - 10^{-1})~\mu m$ in diameter. 
They catalyse the formation of clouds by providing a surface for water vapour to condense upon. These particles of about one-hundredth the size of a cloud droplet are absolutely necessary for turning atmospheric water vapour into liquid water droplets at the 
so-called lifting condensation level, as condensation cannot take place without them due to the Kelvin or curvature 
effect. This effect arises due to the fact that vapour pressure over a curved liquid surface, like a droplet, is exponentially higher than over a flat surface \cite{Kelvin,Helmholtz}. This causes smaller droplets to evaporate faster than larger ones, as the surface curvature reduces molecular bonding, making it crucial for cloud formation.

A great deal about the specific conditions for cloud formation has been learned due to the Cosmics Leaving Outdoor Droplets (CLOUD) experiment at CERN that began operations in 2009 (see e.g. Refs. \cite{CLOUD1,CLOUD2}). CLOUD uses an ultra-clean 3-meter stainless steel chamber where temperature and relative humidity can be set to extreme conditions (down to \(-60^{\circ }C\)). The goal was to study the role of aerosols and the link between cosmic rays and cloud condensation. The collaboration comprises an international team of atmospheric physicists, chemists, cosmic-ray and particle physicists. The experiment itself consists of an advanced cloud chamber equipped with a wide range of external detectors to monitor and analyse the scattering results. The atmosphere conditions can be modeled  within the chamber via the 
control of the relevant parameters. The ``cosmic rays" in the chamber are generated by the Proton Synchrotron (PS), a key element in CERN's accelerator complex\footnote{The PS has a circumference of 628 metres and is composed of about three hundred conventional (room-temperature) electromagnets and 100 dipoles to bend the beams round the ring. It allows for accelerating a beam up to $26~GeV$ and in addition to protons, it can accelerate $\alpha$-particles (helium nuclei), oxygen, sulphur, argon, xenon and lead nuclei, electrons, positrons and antiprotons. CERN's PS yields pulses of $400~ms$ and a top flux of $2 \times 10^{13}$ protons $~s^{-1}m^{-2}$.}. 

After detailed simulations that mimic ground level and stratosphere conditions, CLOUD allowed for acquiring in depth knowledge about the role of aerosols and cosmic rays in the formation of clouds and their effect on climate. In broad terms, the main findings can be summarised as follows: 

\noindent
i) Biogenic aerosols emitted from trees, are major sources of cloud seeds, being presumably, the main contributor of the CCN before the Industrial Revolution;  

\noindent
ii) $HNO_3$ and $NH_3$ drive rapid CCN growth in cities, and are responsible of the thick winter smog;
 
\noindent
iii) Iodine particles from sea ice/ocean are a powerful source of Arctic aerosols;

\noindent
iv) Ions from cosmic rays can significantly enhance aerosol formation. Cosmic rays affect Earth's atmosphere by ionising its gases, which can facilitate the formation of aerosol particles that serve as cloud seeds.
CERN's CLOUD experiment, shows these particles can increase aerosol nucleation rates, but their overall impact on cloud cover is limited.

There exists an overwhelming consensus in the scientific community that global warming is being driven by the sharp increase of anthropogenic greenhouse gases in the atmosphere after 1950s. This consensus arises out of the abundance of evidence of the 
anthropogenic causes and out of the failure to account for the observations, as shown by systematic and thorough studies, through alternative causes. Given that the solar irradiance is a major driver of climate and that the solar cycle has an obvious impact on the space weather through its effect on solar wind and cosmic ray flux on Earth, a putative relationship with climate change was suggested \cite{Svensmark}. 
It was argued that cosmic ray ionisation would affect the rate of nucleation of CCN and thus of cloud formation, which would have the net effect of reflecting incoming short wavelengths and trapping outgoing long wavelength radiation. However, 
after CLOUD results, this connection was shown not to be enough to account for the current  warming of the planet \cite{Ormes}. Thus, even though the statement that more cosmic rays leads to more clouds being correct, it does not imply in a significant effect on the observed Earth's warming (see Ref. \cite{Ormes} and references therein). 

Nevertheless, even though the effect of cosmic rays on the cloud formation is not a determinant factor in global warming, the CLOUD experiment shows that it does have a significant impact. This fact motivates our interest in considering an airborne particle accelerator to boost the condensation of low-altitude clouds, the ones that are relevant for reflecting solar radiation back to space, and to precipitate high-altitude clouds that trap solar radiation reflected from the ground.   

The present proposal can be thought as a tool of geoengineering, which are most often controversial. Indeed, some geoengineering proposals, such as for instance, ocean fertilisation and alkalinity enhancement, and, most particularly, stratospheric aerosol injection (SAI) \cite{Budyko,Crutzen,Rasch,Lenton}, have harmful or, at best, unknown side effects. However, some other proposals are more benign. These include albedo enhancement through passive daytime radiative cooling \cite{Zeven1,Wang}, use of sky-facing thermally-emissive surfaces to radiate heat back into space \cite{Chen,Munday}, cloud brightening or even a large set of mirrors in the sky to reflect back into space a fraction of the incoming solar radiation (see Ref. \cite{Lenton} for a review).  Space reflectors such as a space mirror \cite{Early,Roy,Angel} or a myriad of reflecting bubbles have also been suggested \cite{MIT}. 

It has also been suggested that $CO_2$ could be ejected into space by some suitable adaptations of the much discussed space lift. It was shown that the well of a geostationary orbital lift or space elevator as it is usually referred to, could be used for dumping greenhouse gases into space \cite{OB2023}. Naturally, it was assumed that the stringent requirements to build a stable orbital lift are satisfied and it was discussed how to use this infrastructure for dumping greenhouse gases away from Earth's atmosphere.
In Ref. \cite{OB_Matos}, it was suggested that reflective vessels filled with $CO_2$ at the space end of the lift, could provide an all spectrum shade. It was also considered the case of refractive $CO_2$ filled vessels, which could provide an infrared ``shadow" due to the net effect of attenuating the incoming solar radiation in the $4.3~\mu m$ and $15~\mu m$ wavelengths \cite{Wei}. The use of other greenhouse gases, such as for instance methane, could extend the ``blanket" effect to other wavelengths in the infrared. 
These and some other proposals involving space have been recently discussed in Ref. \cite{Space}.

Getting back to the present proposal, we stress that it arises out of the growing necessity of urgent measures to mitigate the effect of the continuous climbing of the concentration of greenhouse gases, which is dangerously driving the Earth System (ES) towards a Hothouse Earth State \cite{Steffen2018}. 
The proposal might allow for buying some time till a comprehensive substitution of the fossil fuels is implemented and the Earth System recovers. It should be pointed out that the most pessimistic predictions about a possible collapse of the great regulatory ecosystems discussed in Ref.  \cite{Steffen2018} are in agreement with theoretical analyses based on a thermodynamic model of the Earth System and the Hothouse Earth State as an inevitable outcome given the present intensity 
of human activities (see e.g. Refs. \cite{OB-FF1,OB-FF2,OB-FF3,OB-FF4,AB-OB,OB-FF5,OBE}). In the context of this thermodynamical model, building up resilience was shown to be theoretically achievable through a control of the Planetary 
Boundaries and its interactions \cite{OBN}.

Thus, in this context of urgency, we argue that it would be interesting to consider an airborne proton accelerator to enhance the formation of low-altitude clouds for boosting the overall Earth's albedo  and to have some control on precipitation of 
high-altitude clouds. In what follows we shall discuss some benchmark values concerning the present proposal.


\noindent
2. Airborne particle beams have been considered in 1980s as a potential weapon aimed to cripple or destroy incoming intercontinental ballistic missiles (ICBMs). At that time they were conceived as part of the so-called Strategic Defence Initiative.  
Their general principals were discussed in Ref. \cite{RR} and the BEAR 
(Beam Experiment Aboard Rocket) experiment \cite{BEAR}, devised in close collaboration with Los Alamos National Laboratory and industrial partners, presented an example of a successful launch and operation of a linear accelerator in a rocket, 
on July 13th, 1989. 
The flight has shown that a neutral hydrogen beam could be successfully propagated in an exoatmospheric environment. The accelerator produced a $10~mA$, $1~MeV$ beam in $50~\mu$s pulses and 
operated during a few minutes. The major components 
of the accelerator were a $30~KeV$ H-injector, a $1~MeV$ radiofrequency quadrupole, two $425~MHz$ RF amplifiers, a gas cell neutraliser, beam optics, vacuum system and control systems. The device survived the hardships of launch and operated 
autonomously. The robustness of the setup was further demonstrated as the accelerator was
recovered after flight and operated again in laboratory.

\noindent
We shall consider the features of the BEAR experiment as a touching stone example of what has already been achieved and can be reproduced. In order that the airborne accelerator emulates the impact of a beam of cosmic rays, its flux must match their 
flux in the range, say $(1 - 10)~MeV$. At this energy range, the cosmic ray flux is strongly modulated by the solar cycle and consists largely by protons, $\alpha$-particles and small amounts of heavy-ions, electrons, positrons and antiprotons. Galactic cosmic rays at GeV energies are not as strongly coupled with the solar activity. The average flux of solar cosmic rays at about $(1 - 10)~MeV$ is approximately of $10^3 - 10^7$ particles $s^{-1} m^{-2} sr^{-1}$ \cite{NOAA}. A radiofrequency cavity can typically accelerate protons up to $2~MeV$ and weights $1.2$ to $1.5$ tons, hence an airborne platform (airplane, Zeppelin or rocket) can accommodate several of these elements and accelerate a linear beam of particles up to, say, $10~MeV$. We shall consider the $MeV$ range of energies, but as we shall see, the crucial issue is the beam flux of the airborne accelerator. 

\noindent
The processes of catalysing the formation or the precipitation of clouds by ions produced by the collisions of protons to say, $N_2$, the main component of the atmosphere, are separated by a threshold rate of collisions , $R$, given by 
\begin{equation}
R = \Phi \sigma n ~~,   
\label{eq:rate}
\end{equation}
where $\Phi$ is the beam flux, $\sigma$, the proton-$N_2$ cross section and, $n$ the $N_2$ number density, Precipitation requires that the rate of collisions must be high. The exact number depends on various environmental conditions, but a tentative figure of merit is about $10^{8}$ collisions per second per cubic meter \cite{Ma}. If we seek that the airborne proton accelerator flux boosts the formation of low-altitude clouds, but not their precipitation, we demand that $R < 10^{8}$ collisions per second per cubic meter. Taking that  $\sigma \simeq 180~mb$ \cite{MacLeod} and  $n = 2.46 \times 10^{25}~m^{-3}$, obtained from assuming PTN conditions, one gets the bound for the beam flux, namely $\Phi \lsim 2.26 \times 10^{7}~s^{-1}m^{-2}$. This flux matches the one of cosmic rays at the maximum of the solar activity. Furthermore,  the resulting requirement is, in principle, achievable given, for instance, BEAR's beam flux capabilities. Thus, the control of the flux parameters of the accelerator does allow for either boosting the formation of low-altitude clouds and, hence contributing to Earth's albedo, or inducing the precipitation of high-altitude clouds. 

On its hand, the mean free path, $l$, of the proton beam in the atmosphere is given by
\begin{equation}
l = {1 \over \sqrt{2} n \sigma } ~.
\label{eq:free path1}
\end{equation}
From the ideal gas equation of state one can rewrite this equation as
\begin{equation}
l = {k_B T \over \sqrt{2} p \sigma} ~,
\label{eq:free path2}
\end{equation}
where $k_B$ is Boltzmann constant, $p$, the pressure and $T$, the temperature. Thus, under PTN conditions, one obtains: 
\begin{equation}
l = 1.6 \times 10^3~m ~, 
\label{eq:free path3}
\end{equation}
which might suggest that the proton beam can travel a long way from its source before it loses its energy. However, the relevant physical parameter to consider is the energy loss due to ionisation, which is estimated by the Bethe-Bloch formula \cite{Bethe,Bloch}. At the considered energy range, $1~MeV$ of the proton accelerator, one has $\beta^2 \equiv (v/c)^2 \simeq 10^{-3}$, where $v$ is the final velocity of the accelerated proton and $c$ the speed of light, and the non-relativistic Bethe-Bloch formula for a proton should be considered\footnote{For a broader discussion of the relativistic Bethe-Bloch formula and its application for particle detection see, for instance, Ref. \cite{Perkins}.}:
\begin{equation}
{dE \over dx} =  - {4 \pi N \over m_e c^2 \beta^2 } \left({e^2 \over 4 \pi \epsilon_0}\right)^2 ln \left({2 m_e c^2 \beta^2 \over W}\right)~,
\label{eq:Bethe-Bloch}
\end{equation}
where $x$ is the travelled distance, $e$ and $m_e$ are electron's charge and mass, $\epsilon_0$ the vacuum electric permittivity, $W$ the excitation energy, approximately given by $11.7~Z~eV$ \cite{Bloch}, where $Z$ is the atomic number and 
\begin{equation}
N = {N_A Z \rho \over A \mu}~,
\label{eq:Bethe}
\end{equation}
where  $N_A$ is Avogadro's number, $A$ is the relative atomic mass, $\rho$ the density of $N_2$ and $\mu$ the unit of atomic mass.

It is easy to estimate that under PTN conditions:
\begin{equation}
{dE \over dx} =  - 3.4 \times 10^{-3} {MeV \over m}~,
\label{eq:result}
\end{equation}
meaning that a proton from the accelerator will lose its energy in about $300~m$ and the action of the accelerator is rather local and ineffective, for the purpose of catalysing the formation or precipitation of clouds, if installed in a satellite or space station. 
Thus, if the proposed device can fill a volume of about $3 \times 10^{-2}~km^3$ with $1.6 \times10^{10}$ protons, satisfying the condition $\Phi \lsim 2.26 \times 10^{7}~s^{-1}m^{-2}$, it can catalyse the formation of low-altitude clouds. In order to catalyse the precipitation of high-latitude clouds it must yield a higher flux of protons. 

\noindent
Let us close this discussion with some general remarks: 

\noindent
i) The computed accelerator flux is about the same order of magnitude of the cosmic ray flux at the peak of the solar activity at the considered energy range, however, as discussed above, the overall cosmic rays yields neither the warming nor the cooling of the planet. On the hand, the proposed setup can, at least in principle, enhance the formation of low-altitude cloud and the precipitation of high-altitude clouds so to yield a net cooling effect. Furthermore, it can, as mentioned, be used as a versatile tool of weather control;  

\noindent
ii) The examined beam energy is much smaller than the one used in the CLOUD experiment, namely $3.5~GeV$, meaning that a more performing device in terms of acceleration could be more directly guided by the results of that experiment, even though at that energy the flux of cosmic rays is lower;  

\noindent
iii) It is clear that particle acceleration shares with electric propulsion the same physical principles.  On its hand, the most performing electric propulsion systems can accelerate and eject rare gases at a specific impulse, $I_{sp} \equiv v/g < 5000~s$ (see e.g. Ref. \cite{EP}), where $g$ is the acceleration of gravity. Hence, considering $I_{sp} \simeq 3000~s$, it follows that $\beta_{EP} \simeq 10^{-4}$.  Thus, the ionisation energy loss per distance of, for instance, argon atoms is many orders of magnitude greater than the one of protons and hence quite ineffective for the purpose of enhancing formation and precipitation of clouds. 

\noindent
iv) The proposed accelerator could, while in flight, be continuously powered by onboard batteries charged by induction engines suitably fit for this purpose in the airborne platform. The operational features of the accelerator will critically 
depend on the intensity and the duration of its beam. Strategies for  boosting the formation of low-altitude clouds and the precipitation of high-altitude clouds will emerge once these parameters are experimentally established under real operational conditions.  
Of course, the numerical considerations presented above are meant to establish the principle of considering an airborne proton accelerator. Specific situations, away from the PTN conditions and from the assumption that air behaves as an ideal gas, may require a more detailed and involved analysis.


\noindent
3. In this note we have argued that an airborne particle accelerator may be an interesting tool for enhancing the formation of low-orbit clouds, the ones that more significantly contribute to the Earth's albedo due to their capacity to reflect back to space 
incoming short wavelength solar radiation. At high-altitudes, the airborne accelerator can be used to precipitate high-altitude clouds that trap long wavelength radiation reflected from the ground. We have considered the proposed device as an artificial source  of low-energy cosmic rays, which were shown to have a non-negligible role in cloud condensation by CERN's CLOUD experiment. The main advantage of the proposed instrument is that its influence can be adjusted to the altitudes where its action of enhancing cloud formation or precipitation is optimal in what concerns reflecting back and letting through solar radiation into space. In this sense this instrument can be regarded as a tool of geoengineering as well as a device of local weather control, in particular, in providing tailor made cooling relieve of overheated areas and precipitation where needed and granted suitable atmospheric conditions.   

As a matter of principle, it is just natural that geoengineering proposals are regarded with a healthy scepticism  as, most often, they tend to focus on a single aspect of the complex climate change problem, 
overlooking long-term colateral effects that are hard, if not impossible to foresee. However, the present proposal has the advantage of following the well established experimental results of the CERN's CLOUD experiment.
In any case, it is no longer possible to disregard the changes in the  climate patterns due to the accumulation of anthropogenic greenhouse gases in the atmosphere. 
Tackling the disruptive effects of this accumulation is a most urgent issue for our collective future. For sure, on its basis the 
solution for the problem requires concrete changes on the very tenets of the consumption society powered by cheap fossil fuels and on the mistaken assumption that Earth has limitless resources and can absorb an endless amount of waste. 
It is clear that no solution can be reached without a sharp reduction on the emissions of greenhouse gases and radical socio-economic changes. Indeed, a finite planet cannot afford an indefinite period of economical growth without an inevitable collapse. It is increasingly evident that a sustainable future is not possible without a balanced global management of the economic growth  and a drastic reduction in the use of fossil fuels. However, even the prompt implementation of these measures will not have an immediate impact. Therefore, even if geoengineering proposals are not a solution for the core problem of climate change, they might play an important role in getting some extra time to implement adaptation and mitigation measures as well as to restore ecosystems and device enduring socio-economic measures that do not continuously harm the well balanced and subtly tuned Earth System we inherited from the Holocene.  

\vspace{1.0cm}

\noindent
{\bf Acknowledgments~~}

\noindent
This work is partially supported by Project CIVITAS NORTE2030-2024-84. The author would like to thank Ricardo El{\'i}sio and Clovis de Matos for the critical reading of the manuscript.



\bibliographystyle{unsrtnat}

\end{document}